\documentclass[aps, prb, reprint, floatfix, citeautoscript]{revtex4-1}
    \usepackage{amsmath}
    \usepackage{amssymb}
    \usepackage{graphics}
    \usepackage{graphicx}
    \usepackage{color}
    \usepackage[colorlinks=true, citecolor=blue, urlcolor=blue, linkcolor=black]{hyperref}
    \usepackage[capitalize]{cleveref}
    \usepackage[detect-weight=true, detect-family=true, separate-uncertainty=true, multi-part-units=single]{siunitx}

\begin{document}
        \title{Nanoscale Real-Time Detection of Quantum Vortices at Millikelvin Temperatures}

        \author{A.~Guthrie}\email{a.guthrie1@lancaster.ac.uk}
        \author{S.~Kafanov}\email{sergey.kafanov@lancaster.ac.uk}
        \author{M.\,T.~Noble}
        \author{Yu.\,A.~Pashkin}

        \author{G.\,R.~Pickett}
        \author{V.~Tsepelin}
        \affiliation{Department of Physics, Lancaster University, Lancaster, LA1 4YB, United Kingdom}

        \author{A.\,A.~Dorofeev}
        \author{V.\,A.~Krupenin}
        \author{D.\,E.~Presnov}
        \altaffiliation[Also at: ]{D.\,V.\,Skobeltsyn Institute of Nuclear Physics, M.\,V.\,Lomonosov Moscow State University, Moscow, 119991, Russia}
        \affiliation{M.\,V.\,Lomonosov Moscow State University, Quantum Technology Centre, Moscow, 119991, Russia}
        \affiliation{M.\,V.\,Lomonosov Moscow State University, Faculty of Physics, Moscow, 119991, Russia}

        \maketitle

        \textbf{Since we still lack a theory of classical turbulence\,\cite{ClayInstitute}, attention has focused on the conceptually simpler turbulence in quantum fluids.  Can such systems of identical singly-quantized vortices provide a physically accessible ``toy model'' of the classical counterpart?  That said, we have hitherto lacked detectors capable of the real-time, non-invasive probing of the wide range of length scales involved in quantum turbulence.  However, we demonstrate here the real-time detection of quantum vortices by a nanoscale resonant beam in superﬂuid  \textsuperscript{4}He at \SI{10}{\milli\kelvin}.  The basic idea is that we can trap a single vortex along the length of a nanobeam and observe the transitions as a vortex is either trapped or released, which we observe through the shift in the resonant frequency of the beam. With a tuning fork source, we can control the ambient vorticity density and follow its influence on the vortex capture and release rates. But, most important, we show that these devices are capable of probing turbulence on the micron scale.}

        \begin{figure}[b]
            \centering
            \includegraphics[width=\linewidth]{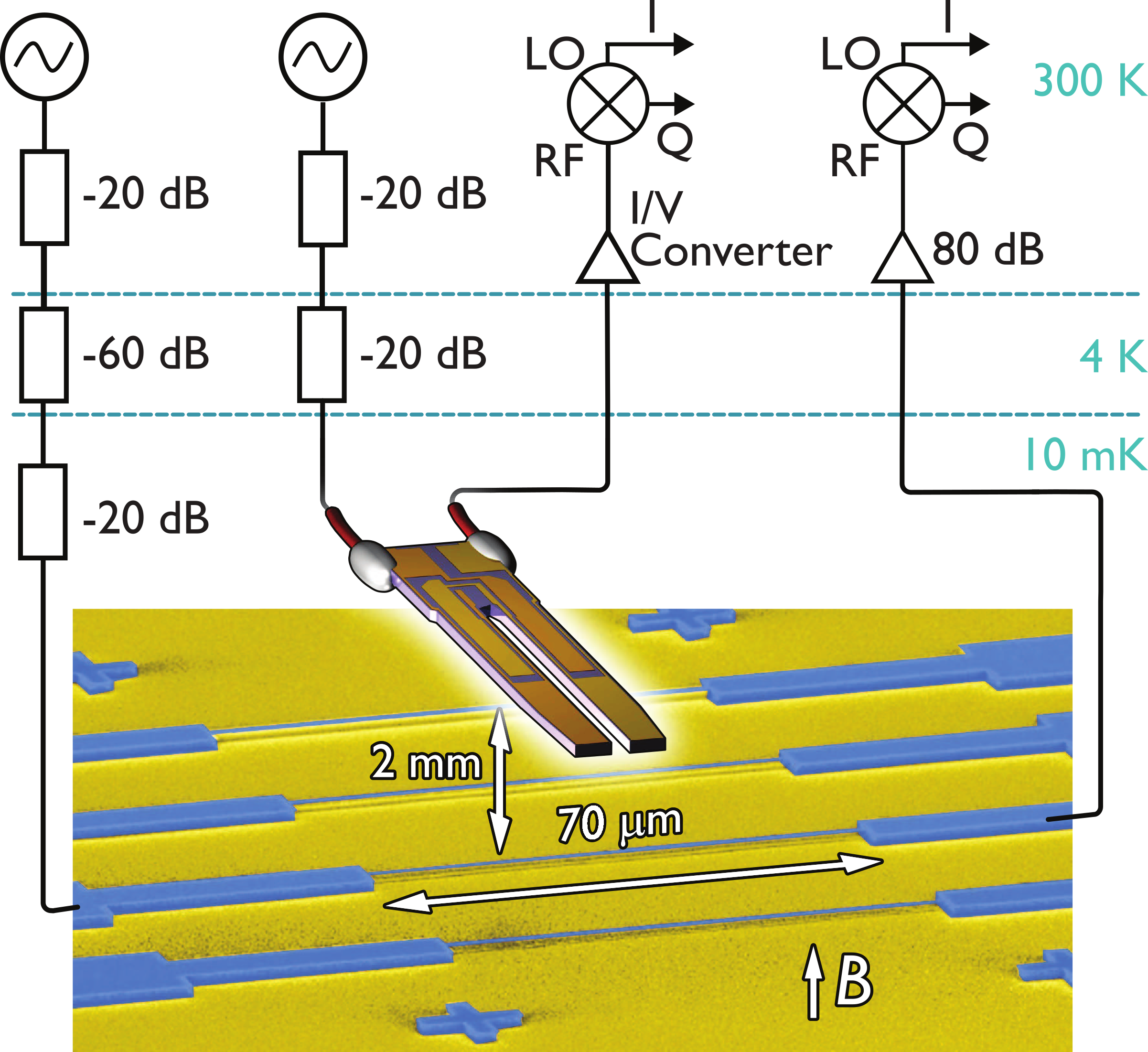}
            \caption{(Colour online) Schematic of the experimental setup.
             A tuning fork generates quantum turbulence, whilst a \SI{70}{\micro\meter}-long nanomechanical beam, suspended \SI{1}{\micro\meter} above the substrate, acts as the detector. The beam and fork are driven by vector network analysers or signal generators through several stages of attenuation at various temperatures. The beam and fork signals are amplified at room temperature by a \SI{80}{\decibel} amplifier and an I/V converter\,\cite{RevSciInstr.83.064703(2012)}. For a detailed description see the Supplementary Information.}
            \label{fig:experiment-schematic}
        \end{figure}

       The nanobeams we use for detecting single vortex events in real time have characteristic dimension less than \SI{1}{\micro\meter} and response times faster than \SI{1}{\milli\second}. Such devices have recently emerged as highly sensitive probes of hydrodynamic\,\cite{SciRep.7.4876(2017), NanoLett.19.3716(2019)} and ballistic \textsuperscript{4}He\,\cite{PhysRevB.100.020506(2019), PhysRevB.101.060503(2020)}.  We present here events demonstrating single-vortex capture, its interaction with the surrounding vortex tangle, and subsequent release via reconnection with a nearby vortex in the surrounding tangle. These measurements advance our capability to probe vortex tangles on much smaller scales than has hitherto been possible.

       \begin{figure*}
            \includegraphics[width=\linewidth]{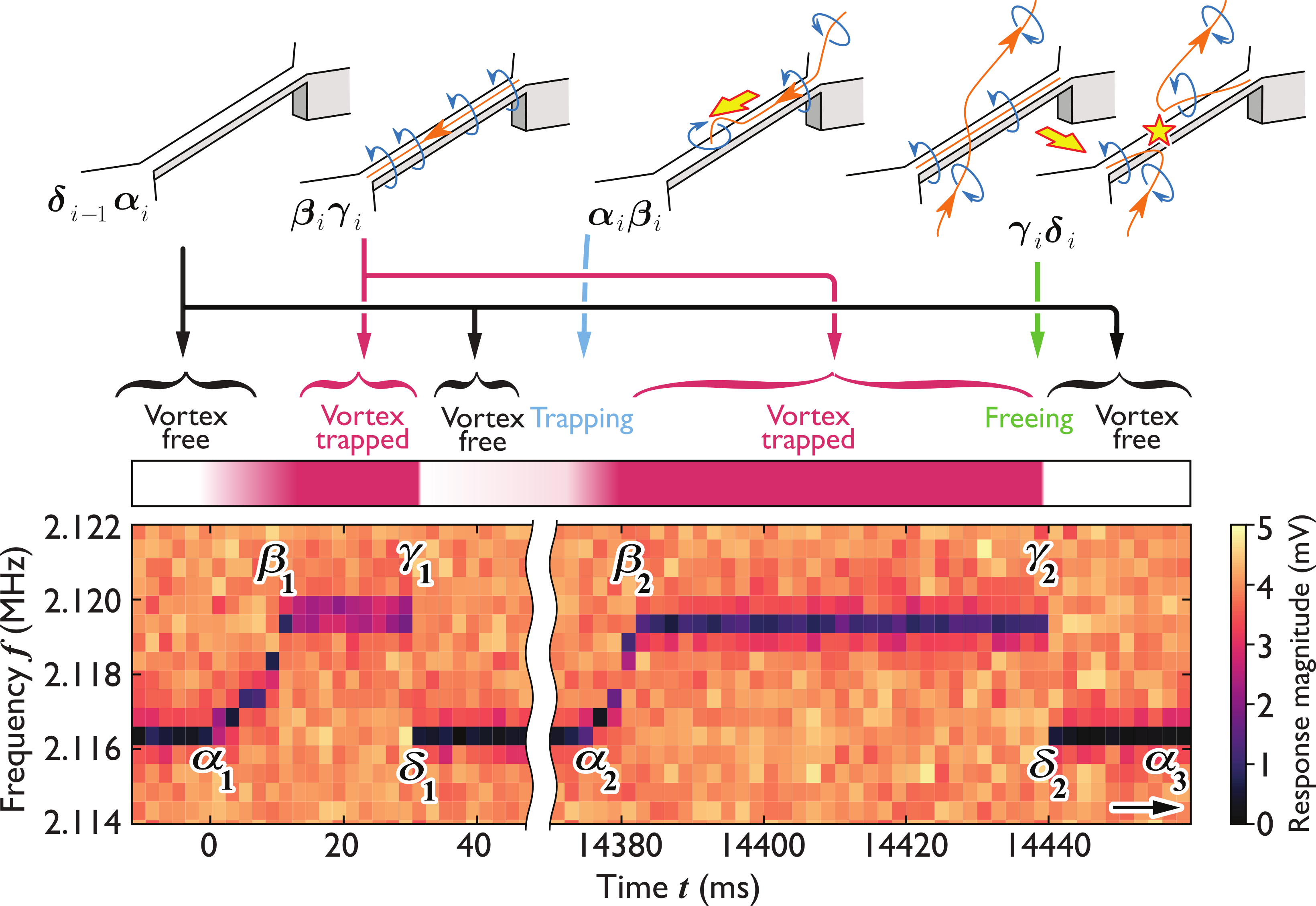}
            \caption{(Colour online) The magnitude of the nanobeam response at each excitation frequency against time taken from the start of the first event in heat-map format. Before point \(\alpha_1\) the beam is in the default vortex-free state. Between \(\alpha_1\) and \(\beta_1\) a vortex interacting with the beam gradually raises the beam frequency by \SI{3}{\kilo\hertz}, finally becoming captured along the entire length of the beam at \(\beta_1\).  From \(\beta_1\) to \(\gamma_1\) the resonance is stable for \SI{20}{\milli\second}.  The captured vortex interacts with a nearby vortex and at point \(\gamma_1/\delta_1\) the system suddenly resets via reconnection of the trapped and attracted vortices and the beam resonance jumps back to the vortex-free state. After \SI{14.35}{s} a second event at \(\alpha_{2}\) occurs with similar features.  The cartoons along the top of the figure sketch the broad processes involved, although the precise details of the capture and release mechanisms are not completely understood.}
            \label{fig:event}
        \end{figure*}

    \Cref{fig:experiment-schematic} shows schematically the measurement setup used for the single-vortex detection. Shown in the lower part of the figure, the doubly-clamped, \SI{70}{\micro\meter}-long, \(\mathrm{Al-Si_3N_4}\) nanobeam with a \(\SI{130}{\nano\meter}\times\SI{200}{\nano\meter}\) cross-section provides  the vortex detector. The beam has a vacuum frequency of \SI{2.166}{\mega\hertz} and is driven at a velocity of only a few millimetres per second. This is orders of magnitude below the expected velocity for the onset of turbulence production\,\cite{JLowtempPhys.153.189(2008)}.  Therefore in all our measurements the beam response is linear and virtually dissipationless.  To provide a controlled source of quantum vortices, we use a quartz tuning fork placed \(\sim \SI{2}{\milli\meter}\) above the beam, driven to a velocity high enough to generate quantum vortices in the ambient superfluid\,\cite{JlowTempPhys.156.116(2009), JLowTempPhys.183.208(2016)}. The details of the operating principles and electrical measurement schemes of the fork and nanobeam are described in the Supplementary Information.

    \Cref{fig:event} shows the time evolution of the response of the nanobeam at each excitation frequency as an example of real-time interactions of the nanobeam with quantum turbulence. The time trace spans two similar consecutive interaction events.  The trace clearly shows that the resonant frequency of the beam shifts significantly (by approximately five widths of the resonance) over a short period of time. We monitor such changes in real-time by the use of a 42-frequency comb produced by a multi-frequency lock-in amplifier\,\cite{RevSciInstr.82.026109(2011), JLowTempPhys.184.1080(2016)}. The \SI{2}{\milli\second} time-analysis interval represents an optimal compromise between fast detection and the frequency resolution of the high-Q resonator.

    The pattern of the events in the figure, with the frequency intermittently jumping from a low to a higher value and back again, is maintained over the many hundreds of such interactions we have recorded.  Initially the beam frequency is low and stable. At time \(\alpha\) (see figure) it gradually increases and stabilises in the region \(\beta\) to \(\gamma\) before abruptly resetting to the initial low-frequency state at time \(\delta\). We can identify and associate each change with the successive stages of the nanobeam's interaction with the vortex tangle.

    \begin{figure*}
        \includegraphics[width=0.9\linewidth]{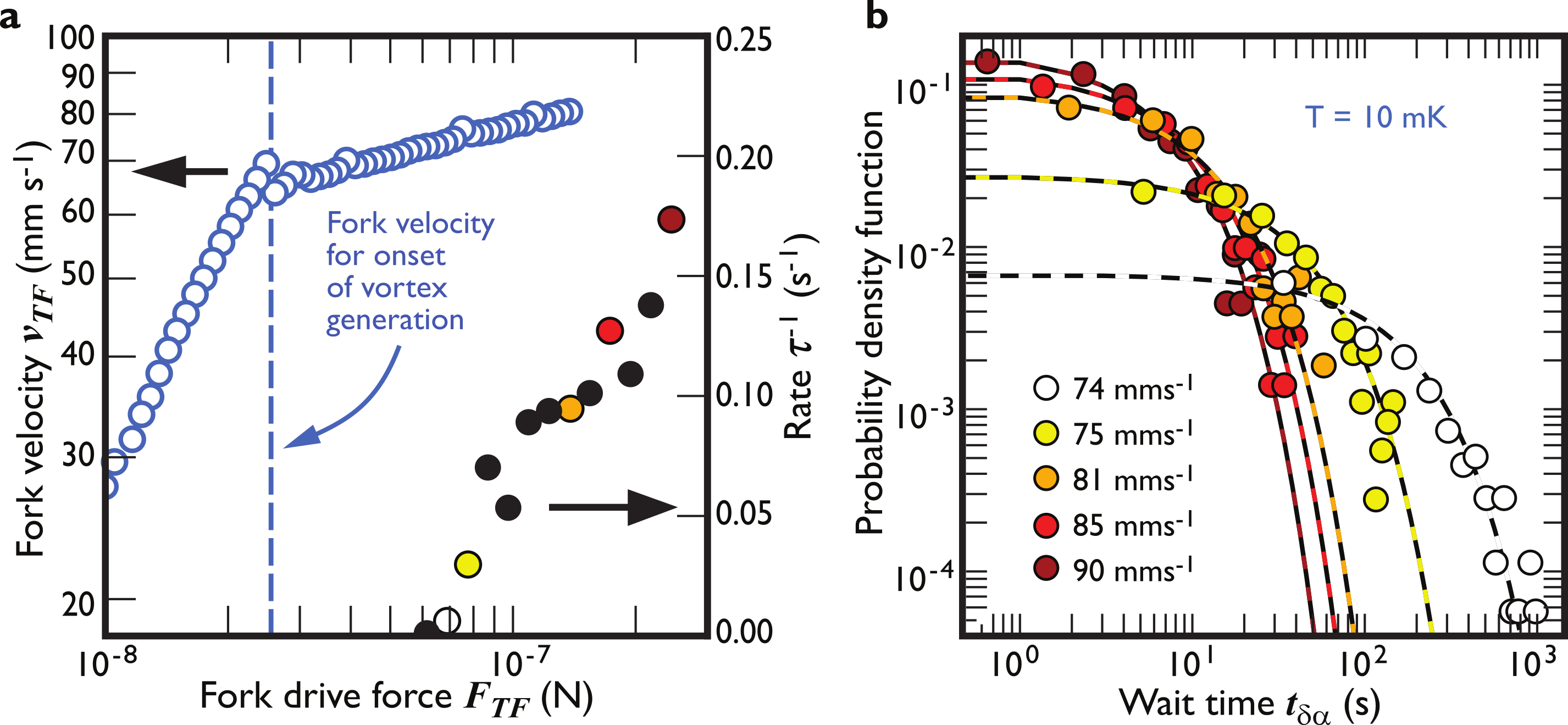}
        \caption{(Colour online) \textbf{a}. The tuning fork velocity as a function of the applied force on the left axis and the rate of detected events by the beam on the right axis. The blue circles correspond to the tuning fork force-velocity dependence, while the symbols on the right show the beam detection rate at various fork forces. The dotted blue line corresponds to the onset of turbulence production by the tuning fork. \textbf{b}. A probability density function of the wait time between events \(t_{\delta\alpha}\) at the same fork velocities.  The solid lines correspond to exponential fits, of the form \(\propto \exp(-t/\tau)\). Note colour and symbol matching between panels \textbf{a} and \textbf{b}. For details, see text.}
        \label{fig:rate-and-fv}
    \end{figure*}

    Referring to \cref{fig:event}, the default state of the beam is that with the lowest frequency (\(\delta_i \alpha_{i+1}\)). This is the only beam response in turbulence-free superfluid, and we identify it with the vortex-free beam. In this state, the beam resonance frequency is reduced by \(\SI{50}{\kilo\hertz}\) from its vacuum value, consistent with the added effective mass contributed by the volume of superfluid displaced.

    The damping of the beam in this state, inferred from the resonance width, is identical with that in vacuum. Therefore there is no significant added dissipation mechanisms in the presence of the superfluid, as expected from the low phonon and roton damping at a temperatures of \(\sim \SI{10}{\milli\kelvin}\)\,\cite{PhysRevB.100.020506(2019)}.

    We believe that the plateau state \(\beta_i \gamma_i\), some \(\sim\SI{3}{\kilo\hertz}\) higher than the low state, represents the case where the nanobeam has trapped a singly quantized vortex along its entire length.  The state is metastable, but will last for several days in the absence of local turbulence, and survive even if the beam motion is ceased, restarted or even driven quite hard. However, upon restarting the turbulence source, the beam relaxes to the default state with the lower frequency described previously.
        
    The identification of the capture of a singly-quantized vortex by the beam is confirmed by several observations. First, the captive vortex generates two additional restoring forces increasing the beam's resonance frequency; one, the force arising from the vortex interacting with its image in the nearby substrate, and two, the Magnus force.

    The interaction of the vortex with a parallel image vortex gives rise to a static force \(\boldsymbol{F} = \rho \boldsymbol{v}_s \times \boldsymbol{\kappa}\), with \(\rho\) the fluid density, \(\boldsymbol{v}_s\) the superflow created by the image at the position of the beam's vortex and \(\boldsymbol{\kappa}\) the circulation\,\cite{ProcRoySocLondA.260.218(1960)}.

    The Magnus force arises from the superfluid circulation around the beam acting on the beam's velocity, \(\boldsymbol{v}\). This force \(\boldsymbol{F}_\mathrm{M}=-\rho \boldsymbol{\kappa} \times \boldsymbol{v}\) causes a periodic displacement of the beam orthogonal to its magnetomotively driven direction.

    While the extra displacement from either of these forces increases the nanobeam's tension, yielding a higher resonance frequency, we find that the image force dominates since the frequency increase depends only weakly on the beam's velocity. For a fuller description of the forces involved see the Supplementary Information.
        
    Secondly, the damping of the beam hardly differs from that of the vortex-free or vacuum state, as expected, since the capture of a single vortex should not significantly change the acoustic emission\,\cite{PhysRevB.100.020506(2019)}, nor should it introduce any new dissipation mechanism. Thirdly, the frequency of the upper plateau is almost always the same (\SI{3}{\kilo\hertz} above the default state), supporting the idea of the capture of a singly-quantized vortex. While double or even higher-order quantization is not energetically unfavourable, it is hard to imagine any creation mechanism. Trapped multiply-quantized vortices would yield discrete higher-frequency plateaus which have not been observed.

    We now can attribute the transitions \(\alpha_i \beta_i\) and \(\gamma_i \delta_i\) between the default and metastable states to the capture and the release of a vortex by the beam. The latter process is always instantaneous on the scale of our detection time and is governed by reconnection of the trapped vortex with crossing vortex in the surrounding superfluid. The dynamics of the vortex capture by the beam is much more challenging to understand and is clearly a more gradual process. In addition to the ``completed’’ events shown in \cref{fig:event}, we also observe many embryonic cases which never fully develop, rapidly reverting to the default state. Here, the implication is that the vortex does not reach the stable state, either from the failure of some intermediate process, or by premature dislodgement by reconnection with a second vortex.

    We should emphasise here that the behaviour of the capture and release processes is completely different.  We can show that by looking at the effect of the local vortex density on these two processes.

    We begin with the effect on the capture process shown in \cref{fig:rate-and-fv}. The local vortex density is controlled by the velocity of the tuning fork\,\cite{JLowTempPhys.183.208(2016)}.  In panel (\textbf{a}) of the figure, we show the tuning fork velocity as a function of the driving force. The clear jump in the slope of the tuning fork response  (marking greatly increased dissipation) indicates  the onset of turbulence production, see for example references\,\cite{JLowtempPhys.138.493(2005), PhysRevB.89.214503(2014)}.

    In panel (\textbf{a}) we also plot a summary of the single frequency measurements of the detection event rate, \(\tau^{-1}\), defined as the inverse mean waiting time \(\tau\) for an event to occur. The event rate increases with the fork's velocity confirming that the nanobeam probes the surrounding tangle density. We only detect vortices at tangle densities corresponding to fork velocities above \SI{73}{\milli\meter\per\second}. At this velocity, the rate of detection is very low with the shortest waiting time between the events being \(\sim\SI{40}{\second}\) and the longest \(\sim\SI{1000}{\second}\). Panel (\textbf{b}) of \cref{fig:rate-and-fv} presents the probability density function (PDF) of the wait time, \(t_{\delta\alpha}\), at five tuning fork velocities showing that the waiting time decreases with increased fork velocity, \textit{i.e.} greater tangle density. The solid lines in the figure correspond to exponential distributions, of the form \(\propto \exp(-t_{\delta\alpha}/\tau)\).  Since it is known that turbulent tangles emit vortex rings following a similar exponential dependence\,\cite{JLowTempPhys.196.184(2019), PhysRevB.101.184515(2020)}, it appears that the capture process may well be governed by the wind of rings emitted by the local tangle. However, whatever the detailed process, it is worth emphasising again that the capture process is governed by the local vortex tangle density.

\begin{figure}
        \centering
        \includegraphics[width=\linewidth]{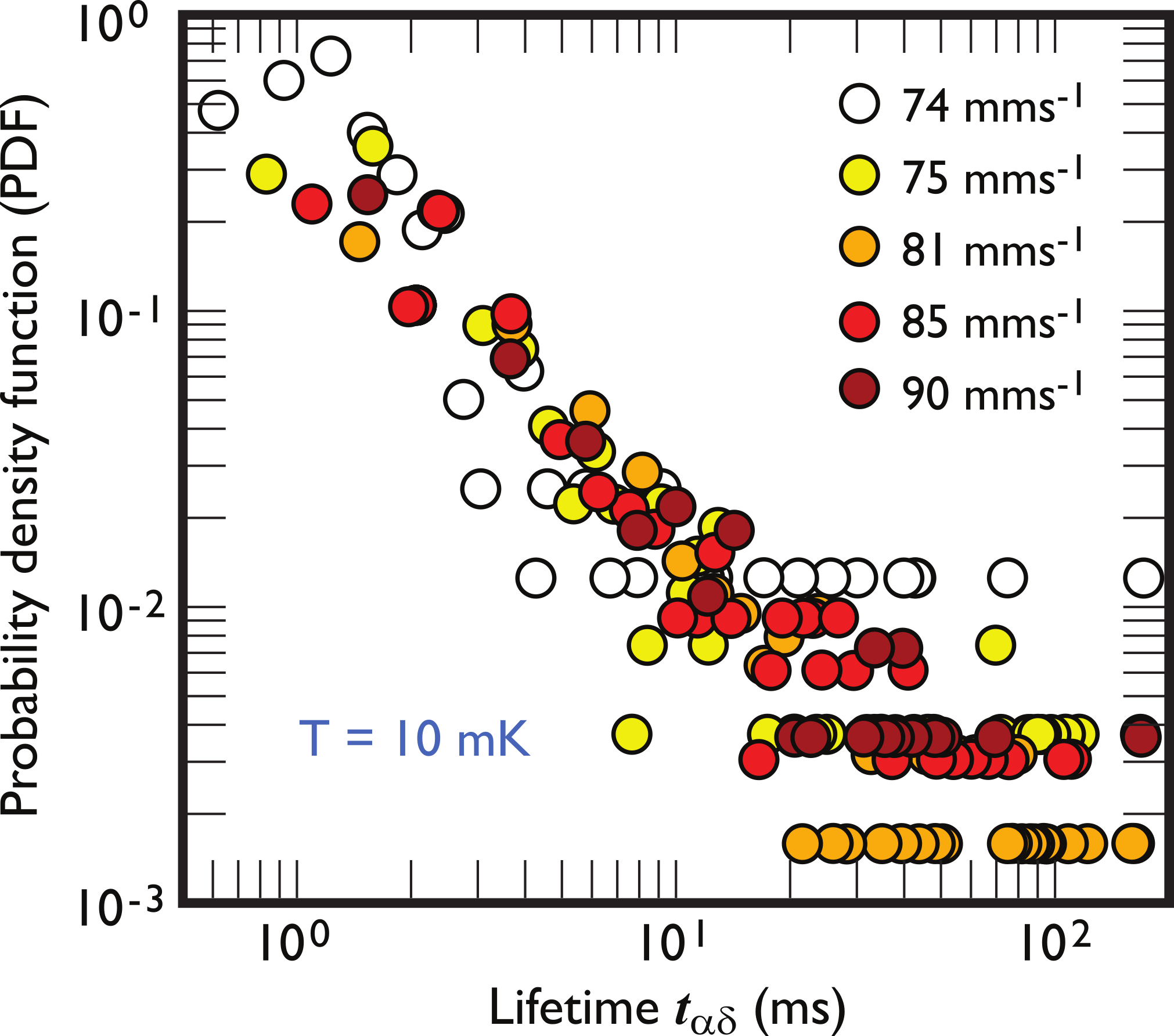}
        \caption{(Colour online) A probability density function (PDF) of captured vortex lifetimes \(t_{\alpha\delta}\) at selected fork velocities. The discrete data at long lifetimes are the result of single observed events.  The data point colours reflect the same data as in \cref{fig:rate-and-fv}.}
        \label{fig:lifetimes}
    \end{figure}

    Once the vortex is captured, its lifetime follows a very different dependence. The release must depend on the proximity of another vortex for annihilation and thus should also carry information on the surrounding tangle. \Cref{fig:lifetimes} shows the probability density function of the measured lifetimes \(t_{\alpha\gamma}\) of vortices on the nanobeam at five tuning fork velocities. First, the typical lifetime of a captured vortex state is three orders of magnitude shorter than the wait time between events. Secondly, the data show no discernible dependence on the tuning fork velocity, showing that the release is insensitive to the overall vortex tangle density. This is surprising since we know that the captured state can exist essentially indefinitely if the vorticity is turned off (carefully to avoid dislodging the vortex in the process). Thus, although we understand the ``on" and ``off" states of the beam, we do not yet fully understand the processes leading to jumps between them.

    Since, in the absence of ambient vorticity, the lifetime of the captured vortex is essentially infinite, the release process must be a result of interaction between the captured and external vortices.  Although the PDF data of the captive lifetime shown in \cref{fig:lifetimes} is too scattered to indicate its functional form, we can use our range of lifetimes to make some rough estimates of the length scales involved. Optical measurements in superfluid helium\,\cite{ProcNatAcadSci.105.13707(2008), ProcNatAcadSci.116.1924(2019)} and simulations of quantum vortex behaviour\,\cite{ProcNatAcadSci.116.12204(2019)} show that the timescale, \(t\), for vortex-vortex interactions displays a square root relationship with the vortex spacing \(\delta\) as \(\delta=A\sqrt{\kappa\,t_{\alpha\gamma}}\), where \(\kappa\) is the circulation quantum and \(A\) a constant of order 1, depending on the geometry of the approaching vortices\,\cite{ProcNatAcadSci.116.1924(2019)}. This expression and our range of lifetimes of 3 to \SI{100}{\milli\second} (as in \cref{fig:lifetimes}), suggests an initial vortex separation of 70 to \SI{230}{\micro\meter}, in excellent agreement, both with typical vortex tangle densities, and the distances reported by the optical measurements.

    In conclusion, we demonstrate that nanobeams can be used as sensitive detectors of single vortex events, tracking their capture, interaction, and release with millisecond resolution, thereby able to probe the local vortex line density. We foresee that we could readily manufacture multiplexed arrays of such beams with the ability to probe the spatial and temporal evolution of a complex vortex tangle with millisecond resolution and potentially single vortex resolution.  Looking further ahead, by capturing a single-vortex in an engineered trapping configuration, we may well be able to study the dynamics of Kelvin waves on the captive vortex, a much anticipated goal in quantum turbulence research\,\cite{JETPLett.111.462(2020)}.

    \textbf{Acknowledgements}
        We thank all members of Lancaster University ULT group as well as A.\,I.~Golov, O.~Kolosov, P.\,V.\,E.~McClintock and W.\,F.~Vinen for helpful discussions. This research was supported by UK EPSRC Grant No. EP/P022197/1 and the EU H2020 European Microkelvin Platform (Grant Agreement 824109). The MSU team was supported by the Russian Science Foundation (Grant 16-12-00072), the research infrastructure of the ''Educational and Methodical Center of Lithography and Microscopy``, M.\,V.\,Lomonosov Moscow State University was used.
    
    \textbf{Authors' contributions} 
        The nanomechanical samples were fabricated by AAD, VAK and DEP. The idea of the experiment was formulated by AG, SK, MTN, YuAP and VT and performed by AG, SK and MTN. The data analysis was done by AG and MTN. The interpretation of the results performed by AG, SK, MTN, YuAP, GRP and VT. The manuscript is mainly written by AG, MTN, GRP and VT.
        
   \textbf{Online content}   
        Statements of data and code availability are available at \url{http://dx.doi.org/10.17635/lancaster/researchdata/xxx}.

		%
		
		\section*{Supplementary Material for: Nanoscale Real-Time Detection of Quantum Vortices at Millikelvin Temperatures}
		
		\section{Device Description}
		The nano-electromechanical device system (NEMS) consists of a doubly-clamped aluminium-on-silicon nitride (\(\mathrm{Al-on-Si_3N_4}\)) composite nanobeam. The beam's dimensions are defined lithographically, with length \(l = \SI{70}{\micro\meter}\) and width \(w= \SI{200}{\nano\meter}\). The \SI{100}{\nano\meter} thick \(\mathrm{Si_3N_4}\) layer determines beam's mechanical properties, while Al layer allows to excite and measure beam motion magnetomotively.  The combined thickness of the aluminium and silicon nitride layers is \(t = \SI{130}{\nano\meter}\), with a combined density of \SI{3062}{\kilo\gram\per\meter\cubed}. The vacuum frequency of the fundamental mode is determined experimentally to be \(f_0 = \SI{2.166}{\mega \hertz}\). The nanobeam is suspended roughly \(d\sim\SI{1}{\micro\meter}\) above the silicon substrate. The experiment is housed in a brass-experimental cell containing superfluid \(\mathrm{^4He}\) at a temperature of \SI{10}{\milli\kelvin}, mounted to the mixing chamber of a cryogen-free dilution refrigerator.
		
		\section{Measurement Scheme}
		The nanobeam response was probed using a magnetomotive detection scheme. Here, the Lorentz force driving the nanobeam is a result of an AC current passed through the nanobeam in a perpendicular magnetic field, which is supplied by a large external solenoid. The beam motion in the magnetic field produces a Faraday voltage that is detected by a drop in the transmitted signal. For characterisation of the nanobeam, a vector network analyser was used to both supply the AC current, and acquire the transmitted response measured as a function of frequency. The resulting Lorentzian resonance curve is fitted to obtain the nanobeam velocity, \(v\), and force, \(F\), using previously established methods \cite{Bradley2017}.
		
		To perform time dependent resonance tracking, two phase-sensitive lock-in measurement techniques were employed: single-frequency detection, and multi-frequency detection. Single-frequency detection was conducted using a signal generator to supply a fixed-frequency, constant AC signal to the nanobeam input, with the nanobeam output connected to a high-frequency (SR844) lock-in amplifier. With the driving frequency fixed on resonance, any change in the nanobeam resonance frequency is detected as a drop in the measured signal.
		
		Simultaneous detection at multiple frequencies was performed using a multi-frequency lock-in amplifier (MLA) \cite{Tholen2011} in place of the signal generator and high-frequency lock-in. The MLA instrument operates by using a frequency comb composed from integer multiples \(n_i\) of a base tone \(f_b\) so that all measurement frequencies \(f_i\) satisfy \(f_i = n_i f_b\). To be able to distinguish between tones the measurement time \(t_m\) must be larger then the inverse separation between frequencies \(t_m > 1/f_b\). This constrains the time resolution and frequency spacing of the instrument, and faster measurements have the frequencies placed further apart. It is also pertinent to note that non-linearity of a resonator will cause mixing between the frequency tones although the use of low excitation drives avoids this problem \cite{Bradley2016}. 
		
		For both resonance tracking techniques, an oscilloscope was used in conjunction with the lock-in demodulation in order to record vortex capture events. The lock-in demodulated signal at the beam's vortex free resonance was monitored by the  oscilloscope, which would trigger the lock-in amplifier to record data when the signal strength fell sufficiently due to resonance frequency shift. For single frequency measurements fall and subsequent rise in the signal would then give the event lifetime. In multi-frequency measurements the recorded data was fitted with a Lorentzian peak to obtain the beam's resonate frequency as a function of time, and the lifetime was then found from this data. 
		
		Similarly, the tuning fork is measured with a vector network analyser, using an I-V converter \cite{Holt2012} (trans-impedance amplifier) to recover the signal which can then be used to find the fork velocity \cite{Karrai1995,Blaauwgeers2007}. The driving force on the fork can be found from the drive signal using well established techniques \cite{Karrai1995,Blaauwgeers2007}. 
		
		\section{Frequency Shift due to a Trapped Vortex}
		\subsection{Tension of the Beam in Vacuum}
		The beam's resonance frequencies can be modelled as the harmonics of a doubly clamped resonator \cite{Bao2005}:
		\begin{equation}\label{eq:frequency}
		f_n=\dfrac{k_n^2}{\pi\sqrt{48}}\dfrac{w}{l^2}\sqrt{\dfrac{E}{\rho_\mathrm{Al}}}\sqrt{1+\gamma_n\left(\dfrac{l}{w}\right)^2 \dfrac{T_0}{wtE}},
		\end{equation}
		where \(\eta\) is the strain, \(w\) and \(l\) represent the width and length of the beam respectively. The coefficients \(k_n\) and \(\gamma_n\) have different values depending on the eigenmode of the resonance: \(k_1=4.7300\), \(\gamma_1=0.2949\), \(k_2 = 7.8532\), \(\gamma_2=0.1453\), and \(k_{n\geqslant 3} = \pi(n + 1/2)\), \(\gamma_{n\geqslant 3}=12(k_n-2)/k_n^3\). 
		
		During fabrication, the \(\mathrm{Si_3N_4}\) layer is pre-stressed to improve the mechanical properties of the beam. Using the measured value of \(f_0 = \SI{2.166}{\mega \hertz}\), with Young's modulus \(E= \SI{70}{\giga\pascal}\) we can estimate the intrinsic nanobeam tension using \cref{eq:frequency} to be \(T_0 = \SI{5.6}{\micro \newton}.\)
		
		\subsection{Hydrodynamic Shift of the Beam Frequency in Liquid \textsuperscript{4}He}
		In liquid \textsuperscript{4}He, the nanobeam fundamental frequency will be shifted due to displacement of fluid by the beam. At \SI{10}{\milli\kelvin} the normal-component is negligible, and we can ignore effects due to hydrodynamic clamping. The hydrodynamic displacement can be modelled as an increase in the effective mass of the beam, which thus shifts the resonance frequency from the vacuum state \cite{Bradley2017}:
		\begin{equation}\label{eq:hydro}
		\left( \frac{f_0}{f_H} \right)^2 = 1 + \beta \frac{\rho_H}{\rho_b}
		\end{equation}
		where \(\rho_H\) is the density of helium and \(\beta\) geometric constant. The resonance frequency for our beam in liquid helium at \SI{10}{\milli\kelvin} is shifted by \SI{50}{\kilo\hertz} from vacuum to \(f_H = \SI{2.116}{\mega \hertz}\). The geometric constant can therefore be calculated as \(\beta = 0.46\).
		
		\subsection{Acoustic damping}
		Acoustic damping is a frequency dependent damping source for oscillators active at all temperatures. However the magnetomotive damping of the beam at \SI{5}{\tesla} is an order of magnitude higher then for acoustic damping \cite{Guenault2019} so we can neglect its affects here. 
		
		\subsection{Effects of a Trapped Vortex on the Beam}
		In order to minimise its energy the trapped vortex will align its core along the nanobeam. The presence of a trapped vortex along the length of the nanobeam will give rise to two forces, both of which act to increase the frequency of the nanobeam. The relative magnitude of these shifts will be estimated here.
		
		\subsubsection{Interactions between the nanobeam, vortex and surface}
		The interaction between the vortex and substrate can be calculated by using the method of an image vortex, \textit{i.e.} one should remove the surface from consideration and assume that the vortex interacts with a parallel image-vortex which is located a distance \(2d\) from it.
		
		The interaction force per unit length between two vortexes is given by:
		\[
		\boldsymbol{f}=\boldsymbol{j}\times\boldsymbol{\kappa},
		\]
		where \(|\boldsymbol{\kappa}|=h/m_\mathrm{^4He}=\SI{9.92e-8}{\meter\squared\per\second}\) is the circulation quanta in \(\mathrm{^4He}\); \(|\boldsymbol{j}|=\rho_{^4\mathrm{He}}v_s\) is the flow density created by image-vortex on the place of the beam; the linear velocity of the superfluid on the distance \(r\) from the vortex core is given by \(v_s=\kappa/(2\pi r)\).
		
		The final expression for the repulsive force between the vortex trapped on the beam and silicon surface is given by
		\[
		|\boldsymbol{F}|=\dfrac{1}{4\pi}\dfrac{l}{d}\kappa^2\rho_\mathrm{^4He}.
		\]
		Substituting our experimental parameters, the force per unit of length and the total attractive force are
		\[|\boldsymbol{f}|=\SI{115}{\nano\newton\per\meter};
		\qquad\qquad |\boldsymbol{F}|=\SI{8.04}{\pico\newton}.\]
		
		Under the action of this force, the beam will sag. Associating an origin with one of the clamped ends of the nanobeam, one can describe such sagging by the function:
		\[
		z(x)=\dfrac{1}{2}\dfrac{|\boldsymbol{f}|}{E}\left(\dfrac{x}{t}\right)^2\dfrac{(l-x)^2}{tw},
		\]
		The maximum displacement at the centre of the beam is:
		\[
		z_\mathrm{max}\left(\dfrac{l}{2}\right)=\dfrac{1}{32}\dfrac{|\boldsymbol{f}|}{E}\dfrac{l^4}{wt^3}
		\]
		in our case the maximum sagging will be \(z_\mathrm{max}=\SI{0.734}{\pico\meter}\).
		
		The maximum tension will be at \(x=0\) and \(x=l\): 
		\[
		T_\mathrm{max} = \dfrac{1}{2}|\boldsymbol{f}|\dfrac{l^2}{t}
		\]
		in our case \(T_\mathrm{max}=\SI{6.51}{\nano\newton}\). The total tension acting on the nanobeam is now given by \(T_\mathrm{tot} = T_0 + T_\mathrm{max}\). Using \(T_\mathrm{tot}\) as the value of the tension, the expected frequency due to a trapped vortex is found by substituting the result of \cref{eq:frequency} into \cref{eq:hydro} as
		\begin{equation}
		f_v = 2.117\,\mathrm{MHz},
		\end{equation}
		and corresponds to a frequency shift of \(f_v - f_H = \SI{1}{\kilo\hertz}\) which is comparable to what was observed.
		
		\subsubsection{Magnus Force on the Beam}
		The value of the Magnus force per unit length in a superfluid liquid is given by:
		\[
		\boldsymbol{r}=-\rho_{^4\mathrm{He}}\,\boldsymbol{\kappa}\times \boldsymbol{v}_b,
		\]
		where \(v_b\) is the beam velocity. Assuming that \(v_b\sim 10^{-2}\,\si{\meter\per\second}\) one can get:
		\[
		|\boldsymbol{r}| = \SI{124.09}{\nano\newton\per\meter},
		\]
		this value is similar to the repulsive force from the surface. However, the sign of the Magnus force depends on the direction of motion. Thus, the half period of the beam oscillations, the Magnus force will be summed with the repulsive force from the surface, while over another half of the period Magnus force will be subtracted from the repulsive force. This leads to a significant diminishing of the observed frequency shift from the Magnus force, observable as a small, velocity dependent contribution to the frequency shift. At a nanobeam velocity of \(v_b\sim10^{-2}\,\si{\meter\per\second}\), the corresponding frequency shift due to the Magnus force was \(\sim \SI{10}{\hertz}\), much less than the contribution from the substrate interaction.
		\providecommand{\noopsort}[1]{}\providecommand{\singleletter}[1]{#1}%
    \end{document}